\documentclass[%
 aps,prl,twocolumn,superscriptaddress,floatfix
]{revtex4-2}

\usepackage{graphicx}
\usepackage{dcolumn}
\usepackage{bm}
\usepackage[utf8]{inputenc}
\usepackage[T1]{fontenc}
\usepackage{mathptmx}
\usepackage{etoolbox}

\usepackage{newtxtext}
\usepackage[varvw]{newtxmath}

\makeatletter
\def\@email#1#2{%
 \endgroup
 \patchcmd{\titleblock@produce}
  {\frontmatter@RRAPformat}
  {\frontmatter@RRAPformat{\produce@RRAP{*#1\href{mailto:#2}{#2}}}\frontmatter@RRAPformat}
  {}{}
}%
\makeatother

\begin{document}

\title{Electrically induced bulk and edge excitations in the fractional quantum Hall regime}

\author{Quentin France}
\affiliation{Department of Physics, Tohoku University, Sendai 980-8578, Japan}
\affiliation{Department of Physics, Sorbonne University, Paris 75005, France}

\author{Yunhyeon Jeong}
\affiliation{Department of Physics, Tohoku University, Sendai 980-8578, Japan}

\author{Akinori Kamiyama}
\affiliation{Department of Physics, Tohoku University, Sendai 980-8578, Japan}

\author{Takaaki Mano}
\affiliation{National Institute for Materials Science, Tsukuba, Ibaraki 305-0047, Japan}

\author{Ken-ichi Sasaki}
\affiliation{NTT Basic Research Laboratories, NTT corporation, 3-1 Morinosato-Wakamiya, Atsugi 243-0198, Japan}

\author{Masahiro Hotta}
\affiliation{Department of Physics, Tohoku University, Sendai 980-8578, Japan}

\author{Go Yusa}
\affiliation{Department of Physics, Tohoku University, Sendai 980-8578, Japan}

\date{\today}

\begin{abstract}

We apply a voltage pulse to electrically excite the incompressible region of a two-dimensional electron liquid in the $\nu=2/3$ fractional quantum Hall state and investigate the collective excitations in both the edge and bulk via photoluminescence spectral energy shifts. Introducing an offset in the voltage pulse significantly enhances the excitation signal. Real-space and time-resolved measurements reveal the dynamics of the bulk excitations, with an estimated group velocity of approximately $3 \times 10^4$ m/s. These bulk excitations align well with the magneto-plasmon model. Our results highlight the topological link between edge and bulk states, providing a novel approach to exploring solid-state analogs of quantum gravity.

\end{abstract}
\maketitle

The quantum Hall effect is a phenomenon observed in a two-dimensional electron gas (2DEG) at low temperatures and under the influence of a strong magnetic field $B$.
In this regime, the longitudinal resistance drops to zero while the Hall resistance becomes quantized.
This occurs when the filling factor $\nu = n_e h / eB$, which represents the ratio of the electron density 
$n_e$ to the magnetic flux quantum density $B/(h/e)$, is either an integer or a rational fraction~\cite{klitzing80,tsui82}. 
Here $h$ and $e$ denote the Planck constant and the elementary charge, respectively. The Hall resistance quantizes to discrete values of $h/\nu e^2$.
The quantum Hall effect represents the first example of topological insulator, characterized by a bulk energy gap and gapless edge~\cite{thouless1982quantized}. 
Bulk excitations typically require energy above the gap, 
which, for the integer quantum Hall effect, corresponds to 
the cyclotron frequency $\omega _{\rm{c}}= eB/m^*$, 
of the electron with an effective mass $m^*$.
In the fractional quantum Hall (FQH) effect, collective neutral excitations
known as magneto-rotons, have been both theoretically predicted \cite{girvin1986magneto} and experimentally observed through Raman scattering \cite{pinczuk1993observation,davies1997spin} and phonon absorption \cite{kukushkin2009dispersion}. 
Recently, magneto-rotons with wavenumbers near zero
have been experimentally studied \cite{liang_evidence_2024},
attracting renewed interest due to their 
anticipated behavior as chiral gravitons with spin 2 \cite{haldane2011geometrical,golkar2016global}.

In contrast to the bulk, the edge
can be excited with infinitesimal energy due to the absence of an energy gap. 
Edge excitations propagate spatially along the edge 
over long distances with suppressed dissipation \cite{ashooriPRB92, kamata2010voltage, matsuuraAPL18,yoshioka2013quantum}.
These edge excitations are characterized by a nearly free propagation  with a group velocity $v_g$
and their frequency is approximately linear 
in the wavevector $k$ as $\omega= v_{\rm{g}} k$.
The unique properties of the edge have inspired 
proposals for innovative applications including quantum energy teleportation
\cite{hotta2009quantum, yusaPRA11, matsuuraAPL18, hottaPRA14} 
and a quantum gravity simulator \cite{hottaPRA14,hottaPRD22,nambu2023analog, yoshimoto2025hawking} 
which models an expanding universe in (1+1) dimensions. 

\begin{figure}
    \centering
    \includegraphics[scale=1.0, pagebox=artbox, clip]{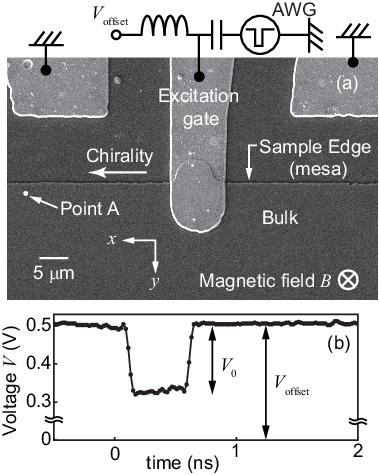} 
    \caption{(a) Scanning electron microscope image of a typical device. The light and dark gray regions correspond to the deposited Au electrodes and GaAs surface, respectively. The excitation gate is connected to an arbitrary waveform generator (AWG) and a DC voltage source, which provides an offset voltage $V_{\rm{offset}}$ to the pulse. The two electrodes connected to the ground and the excitation gate form a coplanar waveguide, minimizing microwave power loss. (b) Voltage waveform applied to the excitation gate, as measured by a $12.5$~GHz bandwidth oscilloscope directly connected to AWG and the DC source. The interval between the measurement points is $20$~ps. $V_{0}$ denotes the amplitude of the square pulse, with a duration of $0.5$~ns.}
\label{fig:fig1}%
\end{figure}

\begin{figure*}
    \centering
    \includegraphics[scale=1.0, pagebox=artbox, clip]{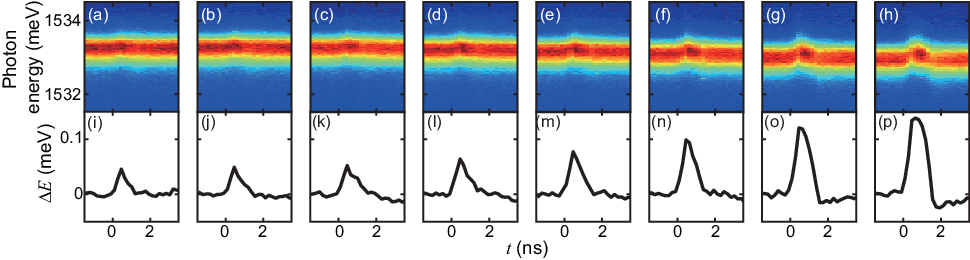} 
    \caption{(a)--(h) Microscopic photoluminescence (PL) spectra as a function of the time delay $t$ between the voltage pulse and the laser pulse for $V_{\rm{offset}}$ ranging from (a) $-0.2$ to (h) $0.5$~V in 0.1-V steps. Measurements were performed at $\nu =2/3$, $B=14$~T and a temperature $T=45$~mK. The measurement point corresponds to the downstream side of the edge (Point A in Fig. 1(a)). (i)--(p) PL peak energy shift, $\Delta E$ as a function of $t$. Each PL spectrum in Figs. 2(a)--(h) was fitted with a Lorentzian function to determine the PL peak energy at each $t$. The energy shift $\Delta E$ is defined as the deviation of the peak energy from the average peak energy observed during the time interval from $t=-1.38$ to $-0.645$~ns prior to the arrival of the edge excitation.} 
    \label{fig:fig2}%
\end{figure*}

The bulk-edge correspondence suggests that both the bulk and the edge 
provide valuable insights into the quantum Hall effect.
We focus on extracting complementary information from the bulk excitations of the $\nu=2/3$ FQH state, 
where the edge modes are known to exhibit considerable complexity. 
We intentionally excited the bulk by efficiently stimulating the edge, using a voltage pulse applied to an electrode connected to both the edge and the bulk. Using a stroboscopic photoluminescence (PL) microscope \cite{kamiyamaPRR22, kamiyama2023dynamics}, we captured the dynamics of these bulk excitations. 
Our results reveal that 
the primary bulk excitations consist of magneto-plasmons, which are
dynamical electromagnetic gauge fields tightly bound to the 2DEG
under an external static magnetic field. 
Magneto-plasmons are intrinsically related to magneto-rotons:
the latters are derived from the lowest Landau level projection of the Hilbert
space~\cite{girvin1986magneto}, while the former can be derived
without this approximation~\cite{macdonald1985magnetoplasmon}.
Furthermore, 
the dispersion relation of magneto-plasmons is expressed as:
\begin{equation}
\omega=\omega_{\rm{c}} + v_{\rm{g}} k.
\end{equation} 
As will be shown later in Eq.~(4), this $v_{\rm{g}}$
has a simple form that is determined solely 
by the quantized Hall resistance
and the permittivity of GaAs.
Additionally, we observed another type of bulk excitation with a velocity approximately an order of magnitude slower than the primary bulk excitation.
This secondary excitation can be interpreted as a strain pulse~\cite{thomsen1986surface}
which represents the bulk counterpart of an edge mode carrying heat \cite{le2022heat}.

The measurements were conducted on a $15$~nm-GaAs/AlGaAs quantum well (QW). An $n+$GaAs substrate served as a back gate electrode, allowing $n_e$ in the QW to be controlled by applying a voltage, $V_{\rm{b}}$, to the back gate. This configuration enables $\nu$ to be set to $2/3$ at various $B$. The horizontal line marked as the sample edge in Fig.~1(a) corresponds to the $\sim$250-nm step of the mesa. The etched region is located on the top side of this setup, while the bulk region containing the 2DEG lies on the bottom side. The excitation gate and the two ground pads, as shown in Fig.~1(a), were fabricated by evaporating Ti/Au over the GaAs surface. Scanning stroboscopic PL microscopy and spectroscopy \cite{kamiyamaPRR22,kamiyama2023dynamics} were used to observe the bulk and edge responses to the incoming AC-pulses transmitted through the excitation gate. A mode-locked Ti:sapphire laser with a central wavelength of $790$~nm and $\sigma \sim 7$ nm ($1.5694$~eV with $\sigma \sim 7$~meV) was used as the light source. The laser emitted pulses are approximately $\sim 1$-ps long with a $13$-ns repetition period. The laser beam was guided to the millikelvin region of the dilution refrigerator via a polarization-maintaining single-mode fiber 
and focused onto the sample using an objective lens. The $\sigma^-$ polarized PL was selectively collected using optics located at the millikelvin region, transmitted through a multi-mode fiber, and subsequently analyzed using a monochromator and CCD detector~\cite{hayakawa}. The measurement point, which corresponds to the focal point of the microscope, was freely movable across the sample using piezoelectric scanners. The time resolution of the stroboscopic PL measurement was approximately $\sim 300$~ps. For more details, refer to the experimental techniques described in \cite{kamiyamaPRR22,kamiyama2023dynamics}. All measurements were conducted at $\nu=2/3$ and a temperature $T$ ranging from $40$~mK to $55$~mK unless otherwise specified. The $\nu=2/3$ FQH state exhibits a competition between the Zeeman and Coulomb energies, leading to a degeneracy between the spin-polarized (ferromagnetic) and unpolarized (non-magnetic) phases at a critical magnetic field $B_{\rm{c}}$. At $B_{\rm{c}}$, a first order phase transition occurs between these two phases \cite{smet2001ising,verdene2007microscopic,hayakawa}, forming domains of these phases and their domain boundaries~\cite{hayakawa}. In our sample, $B_{\rm{c}}$ was determined to be approximately $\sim 7.5$~T based on the $B$-dependence of the PL spectrum \cite{moorePRL17} (see Fig.~S2 in Supplementary Information (SI) for details). A voltage pulse was applied to the excitation gate using an AWG and a DC voltage source, connected via a bias-tee (Fig.~1(a)). This voltage pulse (Fig.~$1$(b)) electrically excites electrons in both the edge and bulk.

To maximize the visibility of the edge excitation, we applied a voltage pulse with an offset $V_{\rm{offset}}$ and investigated its impact on the PL spectrum (Fig.~2). The measurement point (Point A in Fig. 1(a)) was located $x=25$~$\mu$m downstream from the excitation gate and $y=2$~$\mu$m away from the sample edge (mesa). 
PL spectra were recorded by varying the time delay $t$ between the voltage pulse and the laser pulse, with $V_{\rm{offset}}$ as a parameter. Regardless of the $V_{\rm{offset}}$, the PL peak exhibited a distinct blueshift around $t \sim 0.5$~ns compared to the PL peak observed at $t<0$~ns, when the edge excitation has not yet arrived (Figs.~2(a)-(h)) \cite{kamiyamaPRR22}. 

\begin{figure*}
    \centering
    \includegraphics[scale=1.0, pagebox=artbox, clip]{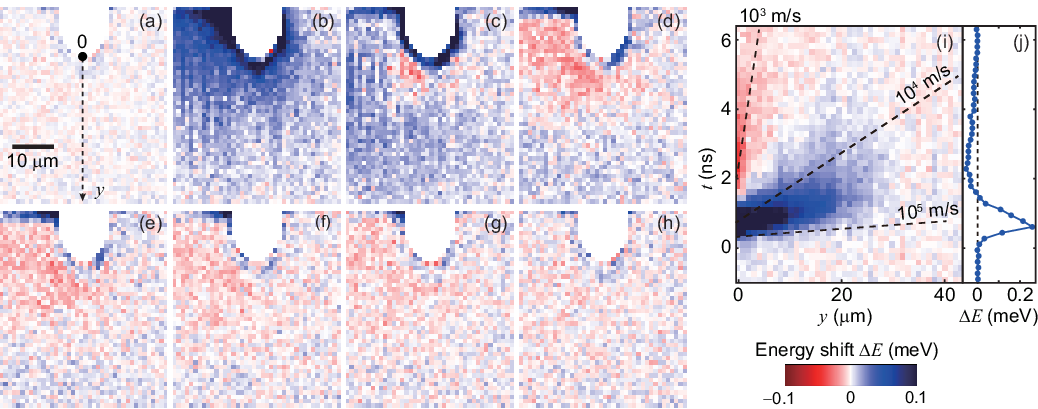} 
    \caption{(a)-(h) Real-space mapping of the energy shift $\Delta E$ captured for several $t$ (a) $t=0.84$, (b) $1.68$, (c) $2.52$, (d) $3.36$, (e) $4.2$, (f), $5.04$, (g) $5.88$, and (h) $6.72$~ns. (See SM movies.) (i) $\Delta E$ dependence both on the spatial position along the $y$-axis (see Fig.~3(a)) and $t$, (j) $\Delta E$ at $y=0$. Note that although all the color maps of Figs. 3(a)-3(h) is unified by $\pm 0.1$~meV for clarity, $\Delta E$ can be more than $0.25$~meV as in Fig. 3(j). All the measurement shown in Fig.~3 was performed at $B=6$~T and $T=45$~mK with $V_{\rm{offset}}=0.5$~V.}
\label{fig:fig3}%
\end{figure*}

To analyze the PL peak energy shift, we fit the PL spectra at a given $(t, x, y)$ using a Lorentzian function
\begin{equation}
L(E; E_{\rm{peak}}, \gamma, A) = \frac{A}{\left(E - E_{\rm{peak}}\right)^2+\gamma^2},
\end{equation}
where $E$, $E_{\rm{peak}}$, $A$, and $\gamma$ represent the photon energy, the central energy of the spectrum, the amplitude, and the full width at half maximum (FWHM), respectively. In this analysis, the fitted $E_{\rm{peak}}$ and $A$ correspond to the estimated PL peak energy and intensity, respectively.
We define the average peak energy
$\tilde{E}_{\rm{peak,0}}$ as the mean of five spectra captured before the arrival of the edge excitation. For Figs.~2(i)--2(p), $\tilde{E}_{\rm{peak,0}}$ 
is calculated by averaging the spectra from $t=-1.38$ to $-0.645$~ns. The PL peak energy shift at a given $(t, x, y)$ is then determined using
\begin{equation}
\Delta E (t,x,y) = E_{\rm{peak}}(t,x,y) - \tilde{E}_{\rm{peak,0}} (x,y).
\end{equation}
Consistent with the PL spectra (Figs.~2(a)--2(h)), $\Delta E$ exhibits a distinct blueshift around $t\sim 0.5$~ns, coinciding with the arrival of the edge excitation at the measurement point. Notably, the maximum value of $\Delta E$ increases as a function of $V_{\rm{offset}}$, reaching more than approximately $\sim 0.1$~meV. Based on this observation, we adopted $V_{\rm{offset}}=0.5$~V for the subsequent experiments discussed below.

By scanning the spatial coordinates $x$ and $y$ around the excitation gate, we captured the $t$-dependence of the PL spectra at $B=6$~T. Since $B<B_{\rm{c}}$, the $\nu=2/3$ state is in the spin-unpolarized phase.
By mapping $\Delta E$ as a function of $x$ and $y$ and using $t$ as the time frame, we reconstructed movies illustrating the propagation of the edge and bulk excitations (see SM for movies recorded at $B=6.5$~T and $11.5$~T).

A single frame of the movie corresponds to the real-space map of $\Delta E$ at a given $t$ (Figs.~3(a)--3(h)). When the voltage pulse is applied to the excitation gate, the PL from the downstream side of the edge exhibits a blueshift (dark blue region in Fig.~3(b)). The edge excitation then propagates along the sample edge (dark blue region in Figs.~3(b)--3(e)). Simultaneously, an excitation near the gate propagates toward the bulk (Figs.~3(b) and 3(c)). Notably, $\Delta E$ near the excitation gate redshifts (red regions in Figs.~3(d)--3(g)), and this negative $\Delta E$ region also spreads into the bulk.

To focus on the behavior of bulk excitations, in Fig.~3(i), we plot the dependence of $\Delta E$ on both $t$ and the $y$ axis perpendicular to the sample edge (see the dotted arrow in Fig.~3(a)). Two distinct modes of bulk excitations are clearly observed, corresponding to the strong blueshift and weak 
redshift regions. The $\Delta E$ at $y=0$ as a function of $t$ exhibits a sharp positive peak at $t\sim 0.5$~ns with an FWHM of $\sim 1$~ns and a negative peak at $t \sim 2-3$~ns with an FWHM of $\sim 2$~ns (Fig.~3(j)). Near the excitation gate, the blueshift exceeds $0.2$~meV. The blueshift region extends to $y\sim 30-40$~$\mu$m, while the redshift region remains relatively closer to the excitation gate at $y\sim 20$~$\mu$m. 

The group velocity $v_{\rm{g}}$ of the bulk excitation is determined by the reciprocal of the slope of the peaks in the $y$-$t$ plot (illustrated by the dotted lines in Fig.~3(j)). To capture this, we measured both the steepest and flattest slopes, corresponding to the earlier and later borders of the blue-shifted region, respectively. Thus, $v_{\rm{g}} \not\equiv l_{\rm p} /\tau$. The group velocity $v_{\rm{g}}$ of the blueshifted peak ranges from $10^4$ to $10^5$~m/s, whereas that of the redshifted peak is on the order of $10^3$~m/s (Fig.~3(j)). Overall, the $v_{\rm{g}}$ of bulk excitations is 1 to 2 orders of magnitude slower than that of the edge excitation \cite{kamiyamaPRR22, kamiyama2023dynamics}. By moving the measurement point further away from the excitation gate, the temporal widths of the blueshift- and redshift modes broaden, indicating that the speed of these modes experience dispersion.

To quantify the behavior of the bulk excitation, we define the penetration length $l_{\rm{P}}$ as the distance along the $y$-direction from $y=0$ to the point where the excitation disappears and the lifetime $\tau$ as the time required for the excitation to vanish. To analyze $l_{\rm{P}}$ and $\tau$, we measured the $y$- and $t$-dependence of the PL spectra at several $B$ ranging from 6 to 10~T (data not shown). By fitting the experimental data at these $B$-values, we obtained $\Delta E$ (see Fig.~S1). From the $y$-$t$ dependence at each $B$, we determined the $B$-dependence of $l_{\rm{P}}$, $v_{\rm{g}}$, and $\tau$, including error estimates (see Fig.~4(a) and SI for details).

\begin{figure}
    \centering
    \includegraphics[scale=1.0, pagebox=artbox, width=\linewidth, clip]{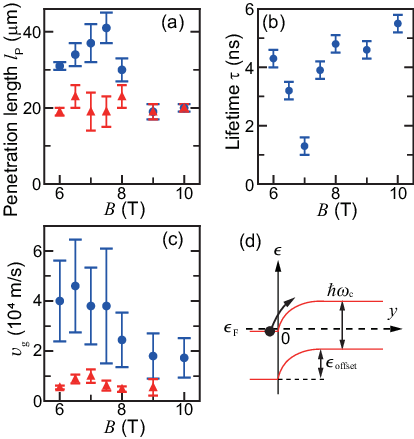}
    \caption{(a) Dependence of penetration length $l_{\rm{P}}$ on $B$.
    The blue and red markers represent bulk excitations where blueshifts and redshifts were observed, respectively. (b) Dependence of lifetime $\tau$ on $B$. 
    (c) Dependence of $v_{\rm{g}}$ on $B$. (d) Schematic illustration of the band diagram of the QW. $\epsilon$ and $\epsilon_{\rm{F}}$ denote the energy of electrons and Fermi energy, respectively. The region $y<0$ corresponds to the area beneath the excitation gate, and $\varepsilon _{\rm{offset}}$ represent the offset energy induced by $V_{\rm{offset}}$ 
    }
\label{fig:fig4}%
\end{figure}

Here, we discuss the mechanism of the bulk excitation at the gate (Fig.~4(d)). 
The distance $d$ between the GaAs surface and the QW is approximately $189$~nm.
The capacitance $C$ of the excitation gate expressed in units of $n_e$ is given by $\varepsilon \varepsilon_0/ed \sim 3.9 \times 10^{11}$ cm$^{-2}$/V (see Fig.~S3). When $V_{\rm{offset}}=0.5$~V is applied to the excitation gate, the electron density under the gate increases by $\Delta n_e\sim2\times 10^{11}$~cm$^{-2}$. Since the density of states of the 2DEG is constant as a function of energy, the energy offset $\epsilon _{\rm{offset}}$ induced by $V_{\rm{offset}}$ can be estimated using $\epsilon _{\rm{offset}}=\pi \hbar ^{2} \Delta n_e /m^*$ \cite{davies1998physics} to be $\epsilon _{\rm{offset}} \sim 7$~meV for $\Delta n_e\sim2\times 10^{11}$~cm$^{-2}$. When the bulk is at $\nu=2/3$ at $B=6$~T ($10$~T), the local electron density under the excitation gate increases to $\sim 3 \times 10^{11}$~cm$^{-2}$ ($\sim 3.6 \times 10^{11}$~cm$^{-2}$) due to $V_{\rm{offset}}$, causing the local filling factor $\nu$ to increase to $\sim 2$ ($\sim 1.5$). Similarly, a pulse amplitude of $V_0=0.2$~V results in an energy change of $\epsilon _0 \sim 3$~meV. 
The electrons near the edge 
can be excited to higher Landau levels because the potential change induced by the voltage pulse (Fig.~1(b)) is of the same order as the Landau level spacing, which is 
$\sim 10$~meV ($17$~meV) at $B=6$~T ($10$~T). 

The dispersion relation of the magneto-plasmon is:
$\omega _{\rm mp} (k) = \sqrt{\omega ^2 _{\rm{c}} + \omega_{\rm{p}}(k)^2}$,
where $\omega_{\rm{p}}(k) = \sqrt{n_e e^2k / 2\varepsilon \varepsilon_0 m^*}$
is the frequency of the surface plasmon 
for the 2DEG. Here, $\varepsilon$ is the relative permittivity of GaAs and $\varepsilon_0$ is the vacuum permittivity~\cite{horing1976quantum,theis1980plasmons,volkov1988edge, macdonald1985magnetoplasmon,sasaki2016determination}.
Since $\omega_{\rm{p}}(k)$ squared is linear in $k$, $v_{\rm g}$ can be derived from the expansion 
$\omega_{\rm mp}(k) = \omega_{\rm c} + \omega_{\rm p}(k)^2/2\omega_{\rm c} + O(k^2)$ as 
\footnote{The wavelength in the horizontal plane, $k^{-1}$, also corresponds to the localization length in the vertical plane. 
We assume that $k^{-1}$ is on the order of a micrometer. For very small values of $k$ that satisfy $\omega_{\rm c} > ck$, the gauge field can radiate out of the 2DEG as Landau emission, causing the magneto-plasmon to become unstable.~\cite{10.1063/5.0233487}}
\begin{equation}
v_{\rm{g}} = \left.\frac{\partial \omega _{\rm{mp}} (k)}{\partial k}\right|_{k \sim 0}
=\frac{\nu e^2}{4 \varepsilon \varepsilon_0 h} 
\left( = \frac{\nu}{2 \varepsilon}\alpha c \right).
\end{equation}
Apart from the environmental factor of $\varepsilon$, $v_{\rm{g}}$
is determined by the fundamental physical constants, similarly to the quantization
of Hall resistivity and depends only on $\nu$. 
Notably, $v_{\rm{g}}$ remains unaffected by changes in $B$, 
provided that the $n_e$ is adjusted to maintain constant $\nu$.
The expression in parenthesis is obtained by introducing the speed of light $c$ 
and the fine-structure constant $\alpha = e^2/2 \varepsilon_0 h c \sim 1/137$. 
Taking $\varepsilon =\varepsilon(\omega)$ 
as the low-frequency limit of the relative permittivity, with 
$\varepsilon(0) = 12.4$ at $\sim 50$~mK \cite{madelung2012semiconductors}, we estimate $v_{\rm{g}} \sim 5.9 \times 10^4$~m/s. This value is consistent with the experimental results, where $v_{\rm{g}} \sim 1$--$6 \times 10^4$~m/s (Fig.~4(c)).

Since magneto-plasmons arise from mixing between adjacent Landau levels,
they provide insights into electron-hole symmetry breaking.
If this mixing is neglected, 
two perspectives on the $\nu = 2/3$ state 
are related by symmetry:
(A) holes form a $\nu=1/3$ state on the fully filled $\nu=1$ ``vacuum'' state of electrons 
(this is known as hole-conjugate picture) and 
(B) electrons form a $\nu=2/3$ state on the empty $\nu=0$ vacuum state. These perspectives,
(A) and (B), are closely linked to edge models~\cite{PhysRevLett.130.076205}.
Magneto-plasmons distinguish these states: the 
$v_{\rm{g}}$ of (A) is half that of (B).
Despite the absence of explicit $B$-dependence in Eq.~(4), 
the experimentally observed $v_{\rm g }$ exhibits slight dependence on $B$ (Fig.~4(c)) which divides the data into two distinct regions: a faster $v_{\rm{g}}$ ($\sim 4 \times 10^4$~m/s) for $B < 7.5$ T and a slower $v_{\rm g }$ ($\sim 2 \times 10^{4}$~m/s) for $B > 8$~T.
The suppression of $v_g$ with increasing $B$
suggests that the hole-conjugate picture (A) 
becomes valid 
when $B > 8$~T.

In addition to $v_{\rm g}$,
there is a possibility that $\ell_{\rm p}$ and $\tau$ are also related to the edge models.
The observed $\tau $ remains relatively constant ($\sim 4$--$5$~ns) 
over a wide range of $B$ except near 7~T (Fig.~4(b))
which is close to $B_{\rm c}=7.5$~T.
At $B \sim B_{\rm c}$, it is known that the spin-polarized (ferromagnetic) and spin-unpolarized (non-magnetic) phases are degenerate \cite{smet2001ising, verdene2007microscopic, hayakawa, moorePRL17}. In this regime, "edge" states form at domain boundaries between different spin phases due to the exchange interaction, which induces an energy barrier at the boundaries \cite{hayakawa, moorePRL17}.
The exchange-energy-induced "edge" is gapless, 
allowing bulk excitations to excite these "edges". 
Consequently, the lifetime of the bulk excitation itself shortens
near $B_{\rm c}$. 
We note that although $\tau$ decreases, $l_{\rm{p}}$ slightly increases at $B_{\rm c}$ (Fig.~4(a)).
This suggests that $l_{\rm p}$ is not solely determined by the magneto-plasmon but also includes contributions from the "edge" at domain boundaries.

The propagation velocity of the redshift mode is on the order of $10^3$ m/s 
and is insensitive to the magnetic field.
These characteristics limit its interpretation. 
This mode is presumably a strain pulse composed of coherent acoustic phonons in GaAs, 
which have a group velocity of $\sim 5\times 10^3$ m/s, \cite{adachi1992}, and the strain induces a redshift of the trion.
It has been reported that such a mode can be excited through the thermal expansion of solids triggered by a picosecond light pulse~\cite{thomsen1986surface}.
In our system, the voltage pulse applied to the gate may induce Joule heating through the magneto-plasmon excitations, which locally increases the temperature and leads to thermal expansion.

In summary, we electrically excited the edge and bulk 
using a voltage pulse and captured images of the bulk excitation dynamics. 
Our findings demonstrate that the primary bulk excitations can be explained by 
magneto-plasmons, while the secondary bulk excitations correspond to a strain pulse. 
Magneto-plasmons are key collective modes that hightlight the topological inseparability of edge and bulk, while strain alters the geometry of the 2DEG.
Our experimental system,
capable of simultaneously observing edge and bulk excitations,
paves the way for future solid-state physics experiments exploring analogous quantum gravity phenomena \cite{hottaPRA14, hottaPRD22, nambu2023analog}. These experiments could include studies of the holographic principle \cite{susskind1995world} 
and bulk-edge correspondence, as exemplified by the AdS/CFT correspondence \cite{maldacena1998, hartman2009cft}.

The authors are grateful to T. Fujisawa, N. Shibata, J. N. Moore, and T. Takayanagi for the fruitful discussions. This work is supported by a Grant-in-Aid for Scientific Research (Grants Nos. 19H05603, 21H05182, 21H05188, and 24H00399) from the Ministry of Education, Culture, Sports, Science, and Technology (MEXT), Japan.

\bibliography{bulk_excitation20250203}

\end{document}